\documentclass[prl,twocolumn,aps,superscriptaddress,letterpaper]{revtex4-1}
\usepackage{amsmath,amssymb}
\usepackage{epsfig}
\usepackage{graphicx}
\usepackage{amsmath}
\usepackage{amsfonts}
\usepackage{epstopdf}
\usepackage{bbold}
\usepackage{color}
\usepackage{microtype}
\usepackage{dsfont}  

\newcommand{\be}{\begin{equation}}
\newcommand{\ee}{\end{equation}}
\newcommand{\bea}{\begin{eqnarray}}
\newcommand{\eea}{\end{eqnarray}}

\newcommand{\nn}{\nonumber\\}
\newcommand{\ba}{\begin{array}}
\newcommand{\ea}{\end{array}}

\newcommand{\R}{{\mathds R}}
\newcommand{\C}{{\mathds C}}
\newcommand{\tS}{\tilde{\mathbb{S}}}

\newcommand{\CO}{{\cal O}} 
\newcommand{\CS}{\mathbb{S}}

\usepackage[colorlinks=true,backref=false, linktocpage=true,
citecolor=blue,urlcolor=blue,linkcolor=blue,pdfpagemode=UseOutlines]{hyperref}

\hypersetup{%
  bookmarksnumbered=true,
  pdftitle = {},
  pdfsubject = {},
  pdfauthor = {},
  pdfkeywords = {}
}

\def\Im {\mathop{\hbox{Im}}}
\def\Tr {\mathop{\hbox{Tr}}}

\begin{document}

\title{Monte Carlo study of real time dynamics}

\author{Andrei Alexandru}
\email{aalexan@gwu.edu }
\affiliation{Department of Physics, The George Washington University,
Washington, DC 20052}
\affiliation{Department of Physics,
University of Maryland, College Park, MD 20742}

\author{G\"ok\c ce Ba\c sar}
\email{gbasar@umd.edu}
\author{Paulo F. Bedaque}
\email{bedaque@umd.edu}
\author{Sohan Vartak}
\email{Sohan@vartak.net}
\author{Neill C. Warrington}
\email{ncwarrin@umd.edu}
\affiliation{Department of Physics,
University of Maryland, College Park, MD 20742}

\date{\today}

\begin{abstract}
Monte Carlo studies involving real time dynamics are severely restricted by the sign problem that emerges from highly oscillatory phase of the path integral. In this letter, we present a new method to compute real time quantities on the lattice using the Schwinger-Keldysh formalism via Monte Carlo simulations. The key idea is to deform the path integration domain to a complex manifold where the phase oscillations are mild and the sign problem is manageable. We use the previously introduced ``contraction algorithm'' to create a Markov chain on this alternative manifold. We substantiate our approach by analyzing the quantum mechanical anharmonic oscillator. Our results are in agreement with the exact ones obtained by diagonalization of the Hamiltonian. The method we introduce is generic and in principle applicable to quantum field theory albeit very slow. We discuss some possible improvements that should speed up the algorithm.  
\end{abstract}


\maketitle

\vskip0.2cm

\section{Introduction}

Except for weakly coupled systems and isolated soluble examples, field theoretical/many-body systems are intractable by analytical means. In those cases numerical Monte Carlo (MC) integration is the method of choice. The computation of equilibrium thermodynamic properties, including equal time correlators, can be recast as the computation of certain well-behaved path integrals, well suited for MC integration, where the integrand decays quickly at large value of the field and is positive everywhere. Some other properties like energy eigenvalues and matrix elements of low lying states can also be recast as  well behaved path integrals  by analytically continuing time to the imaginary direction, effectively using an euclidean space  instead of the original Minkowski space formalism. The success of MC methods in lattice field theory is based on this approach. There are, however, a number of observables that cannot be formulated in this way. They include, for instance, viscosity, conductivity and other transport coefficients \cite{Schafer:2009dj}. They are pervasive in many sub-fields of physics such as heavy ion collisions, neutron star physics, condensed matter and mesoscopic physics and cold atom traps. These observables have in common the fact that they are defined through the thermal equilibrium value of real time (Heisenberg picture) operators of the form
\be\label{eq:corr}
\langle   \CO_1(t) \CO_2(t')   \rangle_\beta = \Tr (e^{-\beta H} \CO_1(t) \CO_2(t')  ).
\ee In thermal equilibrium these correlators depend only on the time difference $t-t'$. Given the obvious importance of these observables, several attempts have been made in the past to compute them with MC techniques. For instance, the complex Langevin method was used in field theoretical models  and in quantum mechanics \cite{Berges:2006xc,Berges:2005yt,Mizutani:2008zz,Fukushima:2015qza}. Despite some early success it seems that the complex Langevin method does not converge when the maximum time difference between operators ($t-t'$) is larger than the inverse temperature $\beta=1/k_BT$. Analytical continuation from Euclidean correlators using techniques such as the Maximum Entropy Method and others has also been studied to compute transport coefficients like electrical conductivity and shear viscosity \cite{Meyer:2007dy,Ding:2010ga,Aarts:2014nba}. Other attempts have also been made in quantum chemistry~\cite{Chang1987,Makri1988,Marchioro1992,Sabo2003}.

\begin{figure}[t]
\includegraphics[scale=0.3]{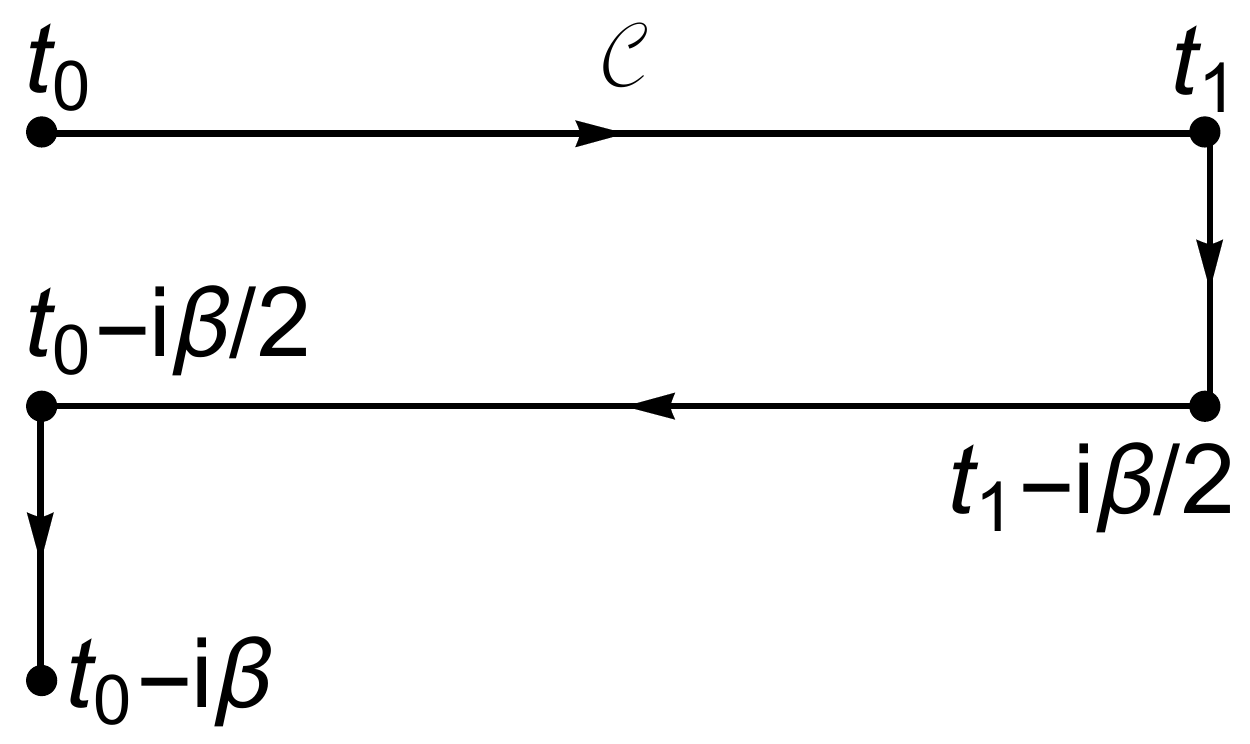}
\includegraphics[scale=0.3]{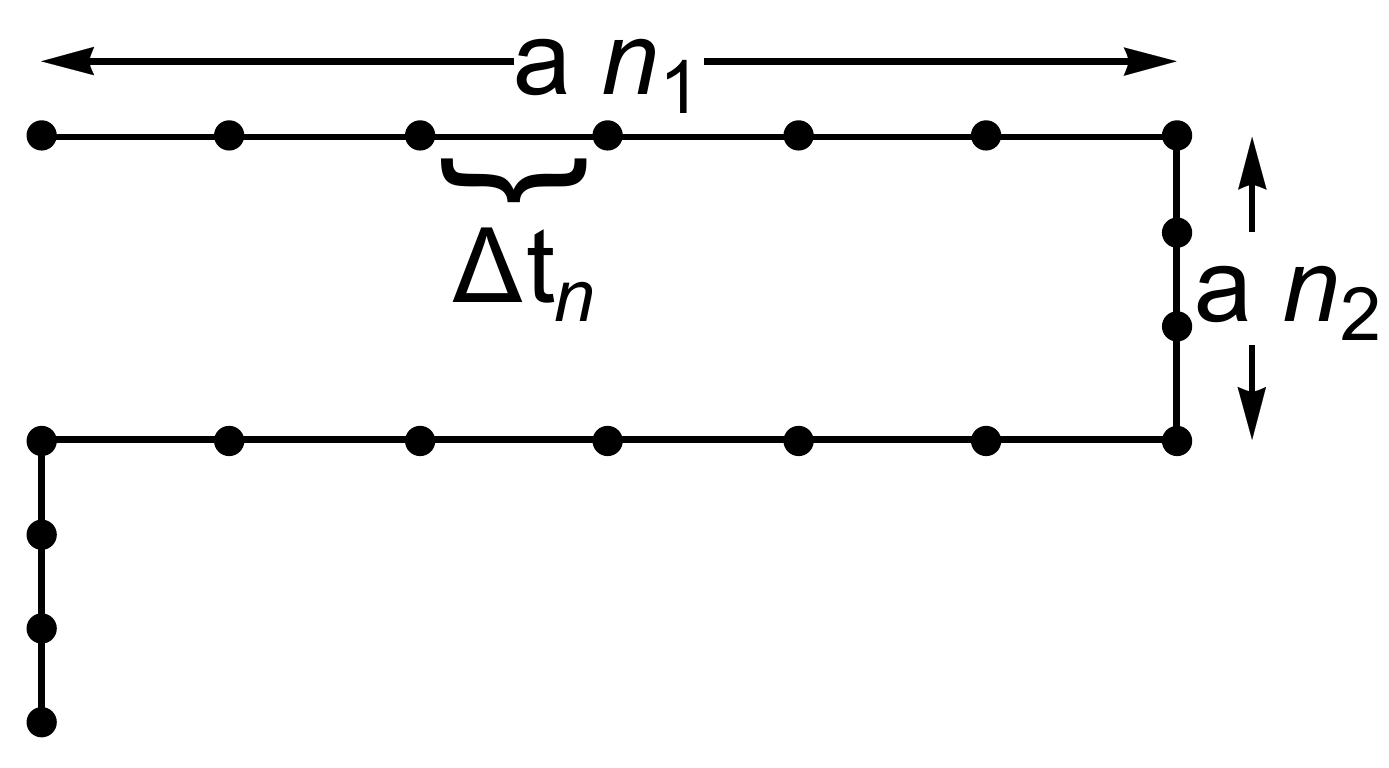}
\caption{The Schwinger-Keldysh contour (left) and its discretized form (right) in the complex time plane. $\Delta t_n$ refers to either $\pm a$ or $-ia$ depending on the location of $n$ on the contour. }
\label{fig:contour}
\end{figure}

Real time correlators of the form shown in \eqref{eq:corr} can be expressed as a path integral using the Schwinger-Keldysh formalism \cite{Schwinger:1960qe,Keldysh:1964ud}. The path integral version of this formalism is summarized in the equations:
\bea\label{eq:SK}
\langle\CO_1(t){\cal O}_2(t^\prime)\rangle &=& \Tr[\CO_1(t)\CO_2(t^\prime)\,e^{-\beta H}]\nn
&=& \Tr[\CO_1(0)\,e^{-iH(t-t^\prime)}\,\CO_2(0) \,e^{iH(t-t^\prime+i\beta) }]\nn
&=& {1\over Z} \int {\cal D}x\, e^{i S_{SK}[x]}\CO_1(t)\CO_2(t^\prime),
\eea 
where $S_{SK}= \int_{\cal C} dt\, L[x]$ is obtained from the original action $S$ by analytically continuing the time $t$ to values on the contour show in Fig. \ref{fig:contour}.~\footnote{There is freedom in choosing this contour. Any contour with initial and final points separated by $-i\beta$ and running through the real axis can be equally used.}

We will describe our  method using the example of a single non-relativistic particle of mass $m$ moving in one dimension under the influence of a potential $V(x)$. The discretized version of the Schwinger-Keldysh action becomes
\be
S_{SK}= \sum_{n=0}^{N} \Delta t_n \left[ {1\over 2} \left( x_{n+1}-x_n\over \Delta t_n \right)^2\!-{V(x_{n+1})+V(x_{n-1})\over 2}\right]
\ee
where $N=2(n_1+n_2)$ (see Fig.~\ref{fig:contour}) and $\Delta t_n$ equals either $a, -a$ or $-ia$ depending whether $t_n$ lies on the positive direction on the real axis, the negative  direction on the real axis or on the segments in the imaginary direction, respectively.

The path integral in the Schwinger-Keldysh formalism poses a tremendous problem for a MC integration since the integrand is very oscillatory leading to subtle cancellations that are hard for the MC method to capture. This difficulty, present whenever the integrand is not positive definite, is known as the ``sign problem". 
The most straightforward approach is to use the reweighting method where $\CS=\CS_R+i\CS_I=-iS_{SK}$ is split into its real and imaginary parts and field configurations distributed according to the (positive) probability distribution $\sim e^{-\CS_R}$ are used to estimate the observable:
\bea
&&\langle   \CO_1(t) \CO_2(t')   \rangle_\beta 
=
\frac{
\int Dx\, e^{-\CS_R} e^{-i\CS_I}  \CO_1(t) \CO_2(t') 
}
{
\int Dx\, e^{-\CS_R} e^{-i\CS_I}  
}\nn
&=&
\frac{
\int Dx\, e^{-\CS_R} e^{-i\CS_I}  \CO_1(t) \CO_2(t') 
}
{
\int Dx\, e^{-\CS_R}  
}
\frac{
\int Dx\, e^{-\CS_R} 
}
{
\int Dx\, e^{-\CS_R}  e^{-i\CS_I}  
}
\nn
&=&
\frac{
\langle   \CO_1(t) \CO_2(t') e^{-i\CS_I}    \rangle_{\CS_R}
}
{
\langle   e^{-i\CS_I}     \rangle_{\CS_R}
}\nn
&\approx&
\frac{
\sum_{a=1}^\mathcal{N} \CO(x_a(t)) \CO(x_a(t'))  e^{-i\CS_I(x_a)}   
}
{
\sum_{a=1}^\mathcal{N}  e^{-i\CS_I(x_a)}   
}
\eea where $x_a(t)$ are a family of $\mathcal{N}$ number of configurations distributed according to the probability distribution $p[x] \sim e^{-\CS[x_a]}$
and $\langle \cdots\rangle_{\CS_R}$ denotes the average computed with the real part of $\CS$ only. The reweigthing method is useful if the average phase $\langle   e^{-i\CS_I}     \rangle_{\CS_R}$ is not too small; otherwise cancellations between the MC estimates of the numerator and denominator lead to large statistical errors. The value of the average phase is actually used as a measure of how hard the sign problem is.
Notice, however, that while $\CS_R$ does not depend on the value of $x(t_n)$ for $t_n$ belonging to the real part of the contour,  $\CS_I$ does depend on $x(t_n)$.  Consequently, the value of $x(t_n)$ (and $\CS_I$) is unconstrained when sampling according to the measure $p\sim e^{-\CS_R}$ and the average phase vanishes identically. Thus, contrary to the usual case where the reweighting method always converges to the correct result, perhaps requiring exponential large number of samples, this method cannot be used in the Schwinger-Keldysh formalism even when infinite statistics is available!

\section{Holomorphic gradient flow}

We attack the sign problem by complexifying the field variables $x_i$. The goal is to replace the original path integration domain, $\R^N$, with an $N$ (real) dimensional manifold embedded into $\C^N \sim \R^{2N}$ such that the variation of $\CS_I$ on this alternative domain  (hence the sign problem)  is milder compared to the one on  $\R^N$ and reweighting can be safely employed. Since the integrands we consider are free of singularities, a multi-dimensional generalization of Cauchy's theorem guarantees that the domain of integration can be changed without altering the value of the integral. The only possible impediment to a deformation of the integration region is the behavior of the integrand at infinity. There are directions in $\C^N$ such that $\CS_R(z)\rightarrow\infty$  as $|z|\rightarrow\infty$. The integral over a domain that asymptotes along these directions is convergent. These ``good'' regions are separated by ``bad'' regions along which $S_R\rightarrow-\infty$ and the integral diverges. Changes in the integration region do not alter the value of the integral as long as the asymptotic behavior is fixed  in one of these ``good" regions at all intermediate steps.~\footnote{Domains of integration are said to be equivalent if the integral over them is the same. The set of \textit{all} equivalence classes is the relative homology group $H_N(\C^N, X)$, where $X$ is the set of disconnected regions in $\C^N$ with $|z|\rightarrow\infty$ and $\CS_R(z)\rightarrow\infty$ \cite{fedoryuk,pham, Witten:2010cx}. 
Crudely speaking $H_N(\C^N, X)$ classifies all the different ways of approaching infinity in $\C^N$ with keeping $e^{-\CS_R}$ bounded so that the integral exists. 
}

We consider a class of manifolds that are generated by the so-called \textit{holomorphic gradient flow} equation,
\bea
\frac{d z_i }{d\tau} =  \overline{  \frac{\partial \CS}{\partial z_i } }\,
\label{eq:flow}
\eea 
(the bar denotes complex conjugation)
which ``flows'' a given point along a curve parameterized by $\tau$ where $\CS_R$ increases the most and $\CS_I$ remains constant.  It is straightforward to show that (i) $d\CS_R/d\tau\geq0$ where the equality only holds if $z_i$ is a critical point (i.e. $d\CS/dz(z_i)=0$), and (ii) $d\CS_I/d\tau=0$. 
Consider the manifold $\Gamma$ obtained by flowing every point $x_i\in \R^N$ by a fixed amount $T_\text{flow}$. 
 The monotonicity property (i) implies that the integral over the manifold $\Gamma$ is the same as over  $\mathds{R}^N$. This is because domains that belong to inequivalent domains of integration are separated by regions where the integral is ill-defined. However, for any $T_\text{flow}$, $e^{-\CS_R[z_i(T_\text{flow})]}\leq e^{-\CS_R[x_i]}$ and the integral is never ill-defined. It is worth to mention that the same property also ensures the absence of the so-called runaway configurations with arbitrarily large negative actions \footnote{In contrast to our method, in the complex Langevin method points over the whole $\mathds{C}^N$ are sampled and regions of arbitrarily negative $S_R$ could be reached.}.
 
 Due to property (i), the flow increases $\CS_R$ and the only regions with significant statistical weight ($\sim e^{-\CS_R}$) originate out of very small regions in $\R^N$. In those small regions $\CS_I$ varies little and therefore, the sign problem is alleviated in $\Gamma$.
In fact, in the limit $T_\text{flow}\rightarrow\infty$, $\Gamma$ becomes the appropriate sum of Lefschetz thimbles (multi-dimensional stationary phase contours) equivalent to $\R^N$ over which $\CS_I$ is  constant. In other words, the flow zooms in on the regions where $\CS_I$ varies slowly, which was exactly the goal that we aimed for to mitigate the sign problem. However, there is a price to pay: the flow might generate multiple regions with nearly constant $\CS_I$, separated by large action barriers which would cause a multimodal distribution hard to sample. The problem then reduces to finding an appropriate value of $T_\text{flow}$ such that, on $\Gamma$, $\CS_I$ varies mildly enough to allow reweighting, yet the potential barriers are not too high so that the configuration space is accurately sampled. This problem depends on the particular model and the parameters involved.

\section{Contraction algorithm}

As we established the domain of integration, $\Gamma$, that is obtained by flowing $\R^N$ by a fixed $T_\text{flow}$ the next step is to generate a Markov chain on $\Gamma$ sampling it according to the distribution $\sim e^{-\CS_R}$. What makes this task challenging is that $\Gamma$ is curved and
it is not obvious how to make proposal lying on $\Gamma$ as
 there is no local way of characterizing $\Gamma$. To overcome this problem in the context of Lefschetz thimbles, the so-called ``contraction algorithm'' was introduced in~\cite{Alexandru:2015xva} and was generalized to manifolds beyond thimbles in~\cite{Alexandru:2015sua}. We will use the same algorithm in our analysis. Let us begin by reviewing it.  

We use the fact that the flow equation \eqref{eq:flow} defines a one-to-one map between each point $z_i\in\Gamma$ and $x_i\in\R^N$ where $z_i:=z_i(T_\text{flow})$ is the solution of \eqref{eq:flow} with the initial condition $z_i(0)=x_i$, and use $x_i$ to parametrize $z_i$. Using this parameterization we can write
\bea
\int_\Gamma \!\!\!Dz \,^{-\CS[z]}\, \CO[z]&=&\int_{\R^N} \!\!\!\!\!\!d^Nx\, \det J\,e^{-\CS[z(x)]} \CO[z(x)]\,.
\eea  
where $J_{ij}=\left({\partial z_i \over \partial x_j} \right) $ is the Jacobian associated with the change of variables from $z_i$ to $x_i$ and $\CO$ represents any observable as, for instance, $\CO_1(t)\CO_2(t')$. $\det J$ is a complex number and accounts for the change in the volume element as well as the orientation of the tangent plane of $\Gamma$ in complex space. The evolution of the Jacobian matrix $J_{ij}$ with the flow is determined by:
\bea\label{eq:flowJ}
 \frac{dJ_{ij}}{d\tau}  &=& \overline{  \frac{\partial^2 S[z]}{\partial z_i \partial z_k}J_{kj}  }\quad,\quad J(0)=\mathbb{1}
\eea
with $z_i(\tau)$ satisfying \eqref{eq:flow}. 
We can then write
\bea\label{eq:contraction}
\langle\CO\rangle&=&
{ \int d^Nx\, e^{-\tS[z(x)]} \CO[z(x)] }\over {\int d^Nx\, e^{-\tS[z(x)]} } \nn
&=& { \int d^Nx\, \det J e^{-\tS_R} e^{-i\tS_I}\CO \over \int d^Nx\,e^{-\tS_R} }{ \int d^Nx\,e^{-\tS_R}  \over \int d^Nx\,e^{-\tS_R} e^{-i\tS_I} }\nn
&=& {\langle e^{-i \, \tS_I }\CO \rangle_{\tS_R}   \over \langle e^{-i \, \tS_I }  \rangle_{\tS_R}},
\eea
where $\tS[x]= \CS[z(x)]-\log\det J$ is the effective action whose real part determines the probability distribution (i.e. $P(x_i)\propto e^{-\tS_R[x]}$). We use a standard Metropolis algorithm to generate samples. In this method, we make all the updates in $\R^N$ and the flow evolution guarantees that the points 
$z_i$ lie on the manifold $\Gamma$, as desired.
 In the last step of \eqref{eq:contraction} we reweighed the phase which involves both the contribution from $S_I[z(x)]$ and $\Im(\log\det J)$. As we discussed earlier, the variation of $\CS_I[z(x)]$ on regions which dominate the integral is mild. We have also found that $\Im(\log\det J)$ fluctuates very weakly on these regions as well. Therefore reweighting the phase does not produce large errors. In a nutshell, the contraction algorithm is a standard Metropolis algorithm in the variables $x_i$ using the effective action $\tS_R[x]$ where the phase $e^{-i\tS_I}$ is reweighted during the computation of the observable.

\begin{figure}[t]
\includegraphics[scale=0.3]{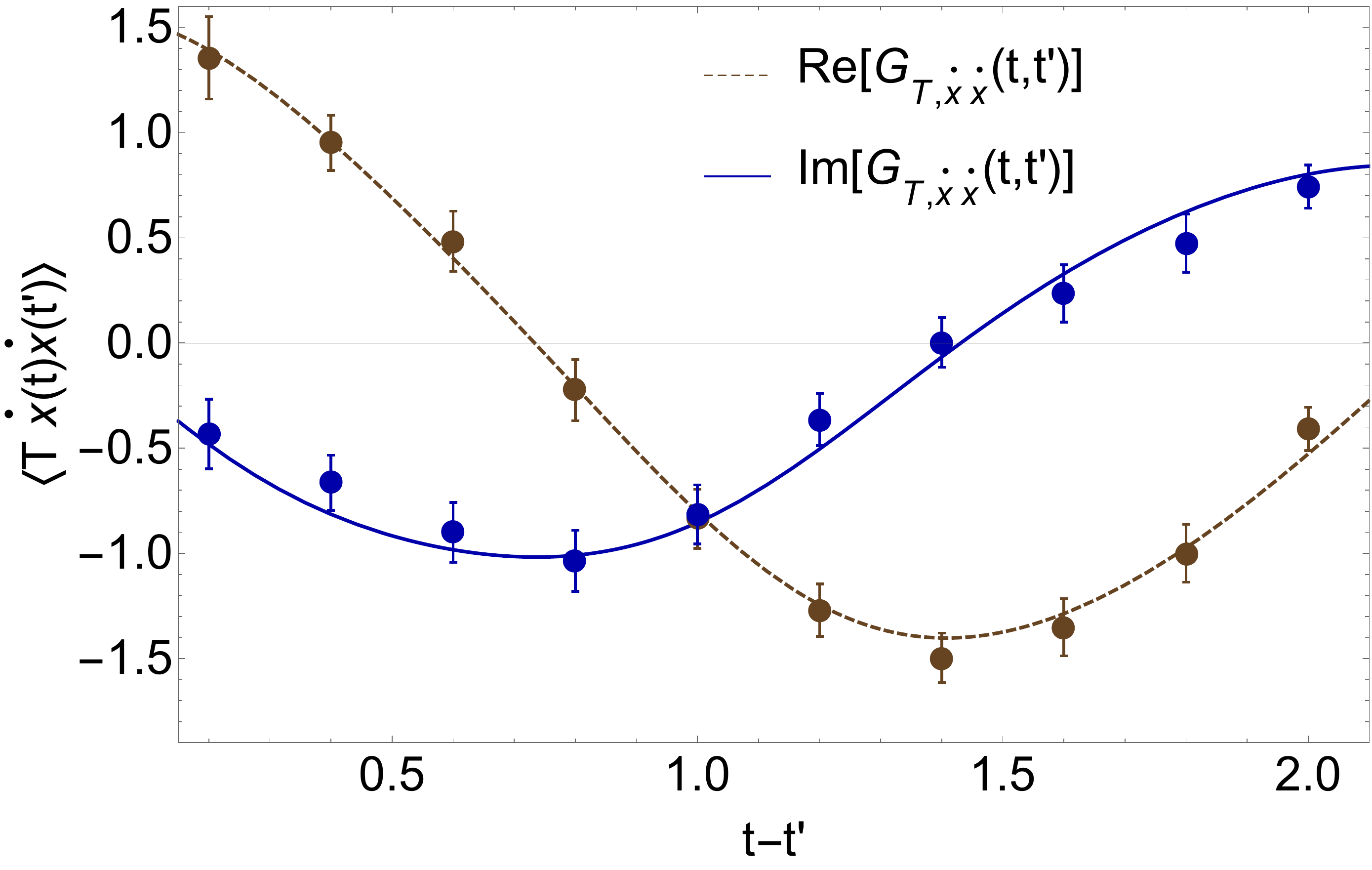}
\includegraphics[scale=0.3]{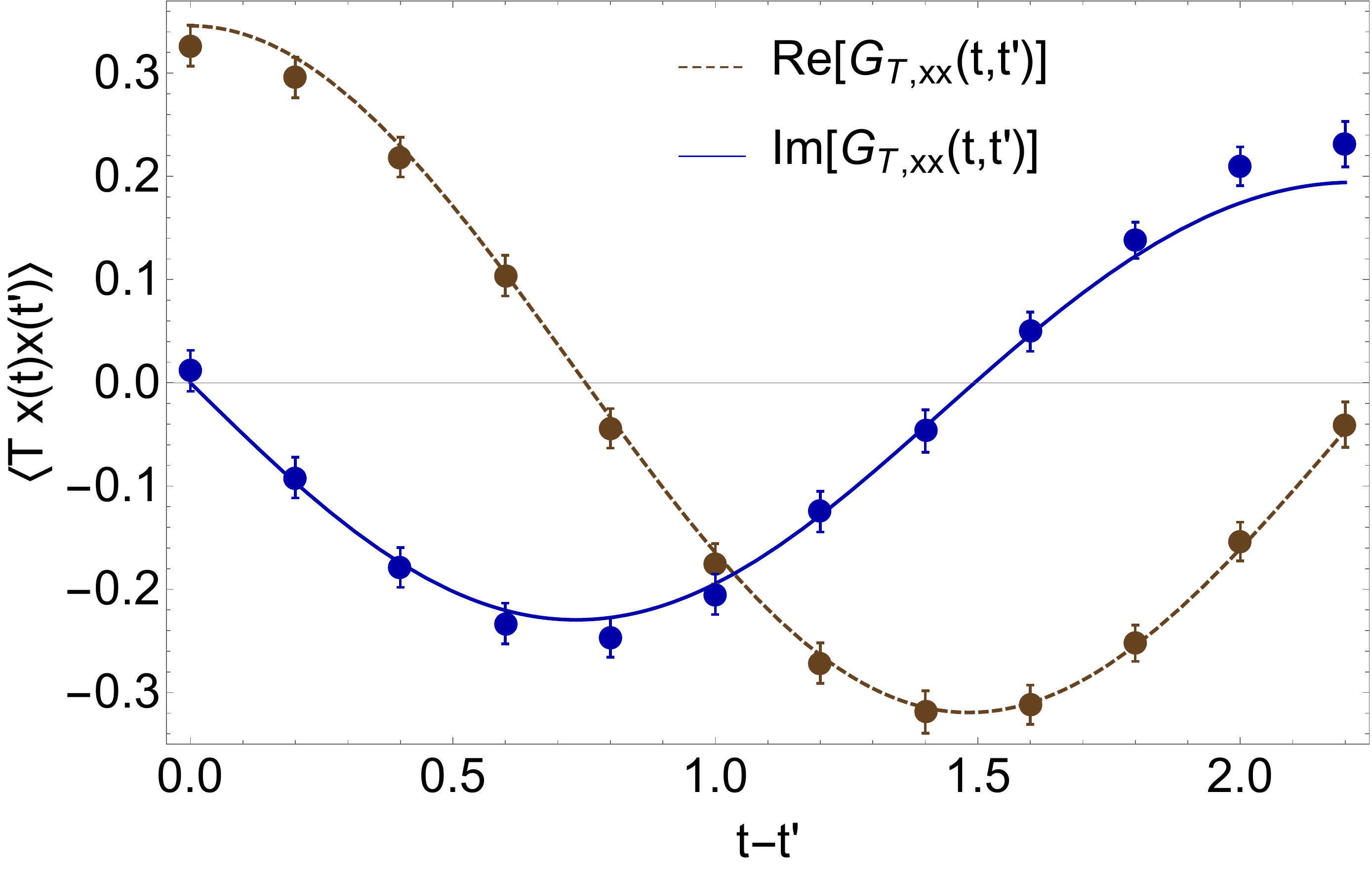}
\caption{Time ordered (Feynman) correlators $\langle T(\dot x(t) \dot x(t^\prime))\rangle$ and $\langle T(x(t)  x(t^\prime))\rangle$. The dotted and solid lines represent the exact results obtained by diagonalizing the Hamiltonian.}
\label{fig:correlatorsR}
\end{figure}

Some care has to be taken regarding the proposals. Due to the nonlinear nature of the flow equation, the image of the directions on $\Gamma$ along which the variation of the action is mild is typically very distorted and anisotropic in $\R^N$, where the updates are made. As a result, there are some steep directions in $\R^N$ along which $\tS[x]$ changes very rapidly and some flat directions where it changes very slowly. We choose the proposals such that the size of the random step is larger along the flat directions and smaller along the steep directions for better efficiency. In order to do so, we use the a quadratic estimate for the effective action $ \tS_R[x]\approx {1\over2} x^T M x$ where $M$ is a real matrix \footnote{The quadratic approximation to $\CS$ is only made in the proposal stage. The sampling is done with respect to the exact $\CS$ without any approximations.  
}. In the quadratic approximation, solution to the flow equation \eqref{eq:flow} hence the matrix $M$ can be found analytically. Any vector $x_i\in \R^N$ can be decomposed in terms of the eigenvectors of $M$ as
\bea
x_i=\sum_{\alpha=1}^N c^{(\alpha)} \rho^{(\alpha)}_i \quad\text{where}\quad M \rho^{(\alpha)}= \lambda^{(\alpha)} \rho^{(\alpha)}\,.
\eea
The eigenvalues $\lambda_a$ provide an estimate for the variation of the action on $\Gamma$ in the direction $\rho^{(\alpha)}$. The proposals are then given by 
\bea
c^{(\alpha)}_\text{proposed}= c^{(\alpha)}_\text{old}+ {\delta \over \sqrt{\lambda^{(\alpha)} }} 
\label{eq:proposal}
\eea
where $\delta$ is a random number satisfying $P(\delta) = P(-\delta)$ to ensure detailed balance. At each update, we randomly select a direction $\alpha$ and propose a step in that direction. The proposed configuration is then accepted with probability $\min\{1, \exp(-\tS[x_\text{proposed}] + \tS[x_\text{old}])\}$.

\section{Results }

We computed the time ordered (Feynman) correlation functions $\theta(t-t^\prime)\langle\dot x(t) \dot x(t^\prime)\rangle+\theta(t^\prime-t)\langle\dot x(t^\prime) \dot x(t)\rangle$ and $\theta(t-t^\prime)\langle x(t)  x(t^\prime)\rangle+\theta(t^\prime-t)\langle x(t^\prime)  x(t)\rangle$ for the quantum anharmonic oscillator with $V(x)={\omega^2\over 2}x^2+{\lambda\over 4!}x^4$. The former characterizes the linear response
of the system to an external force. Of course, since there is a single degree of freedom there is no actual dissipation in our model, but nevertheless this correlator can be thought of a quantum mechanical analogue of conductivity. Our parameters are as follows: lattice spacing $a=0.2$, $n_1=12$, $n_2=2$ (i.e. $t_\text{max}=2.2$, $\beta=0.8$), $\omega=1$, $\lambda=24$ and $T_\text{flow}=0.2$. The choice of the coupling constant $\lambda$ is such that the anharmonic term is of the same order as the quadratic mass term and the theory is in the strongly coupled regime. Our results for the real and imaginary parts of the retarded correlators are plotted in Fig. \ref{fig:correlatorsR}.

The problem we studied has been studied via complex Langevin method in the past~\cite{Berges:2005yt,Berges:2006xc,Mizutani:2008zz}. However one vital shortcoming of the complex Langevin approach is that it does not converge to the correct answer for operators separated by real time intervals $t_\text{max}>\beta$. For the purposes of computing transport coefficients, which are expressed as small frequency limits of the Fourier transforms of the time dependent correlation functions, it is important to be able to accurately compute correlators with $t_\text{max}\gtrsim\beta$. This is  because one expects, for strongly coupled theories,  the damping time to be proportional to $\beta$ and depending on the proportionality constant the main support to the Fourier transform can extend to $t\gtrsim\beta$. For theories with intermediate coupling where the damping time might be greater, the problem gets worse. Remarkably, our approach does not suffer from this problem. In fact, in the computations presented in Fig. \ref{fig:correlatorsR} we were able to go as high as $t_\text{max}\approx 3 \beta$. 

\begin{figure}[t]
\includegraphics[scale=0.3]{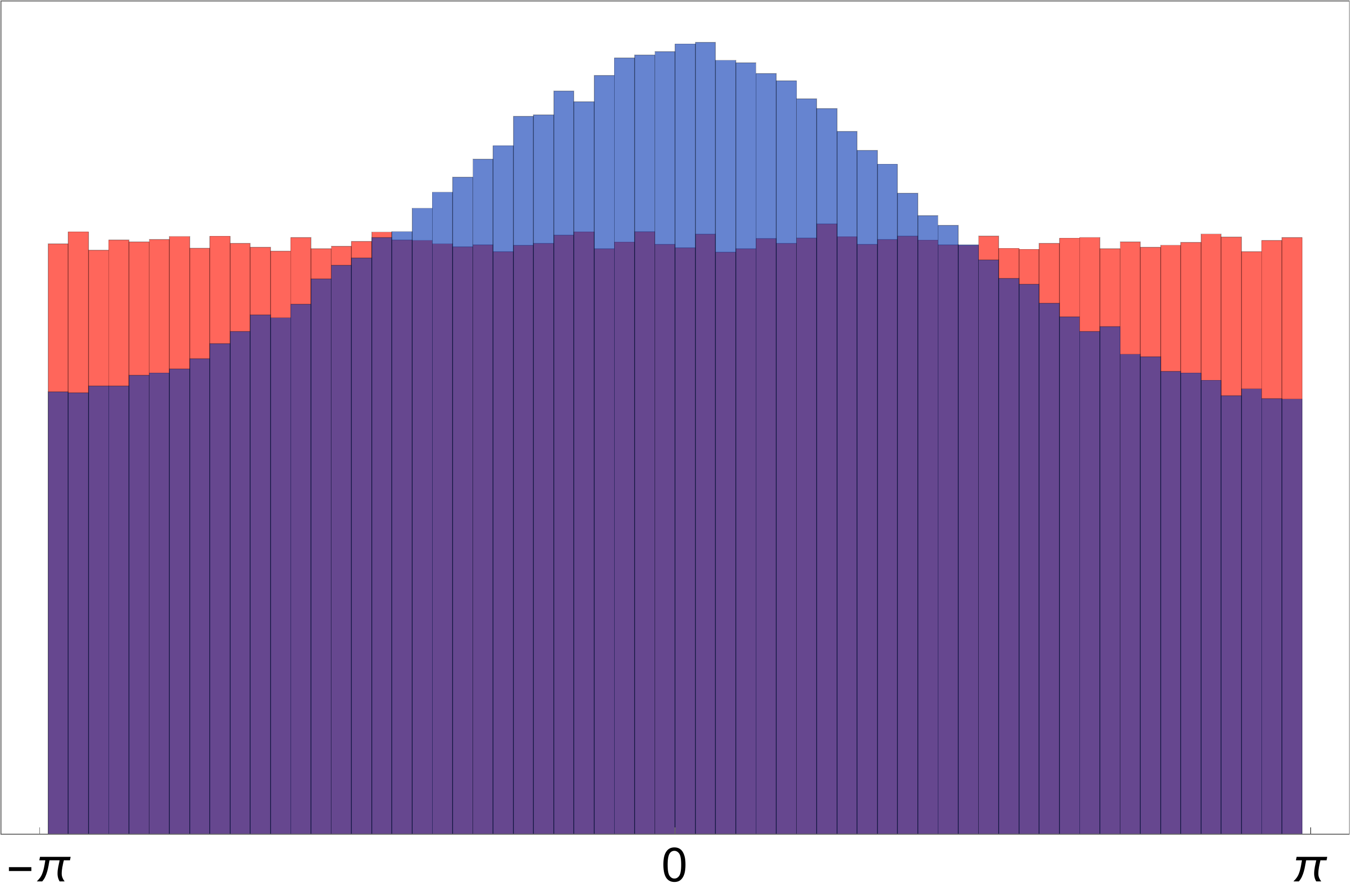}
\caption{Histogram of $\CS_I$ (mod $2\pi$) for the $T_\text{flow}=0$ calculation corresponding to an integration over $\R^N$ (in red) and the $T_\text{flow}=0.2$ calculation corresponding to an integration over $\Gamma$ (in blue). It is clear that the modest flow $T_\text{flow}=0.2$  reduces the sign problem significantly.}
\label{fig:histogram}
\end{figure}

The convergence to the right result is encouraging and, to our knowledge, unique to our method.
There is, however, room for improvement.  
The convergence is rather slow; the results above required $3\times10^7$ Metropolis steps.
Going to high values of  $t_\text{max}$ requires more computational effort. The dependence of the cost on the parameters, such as $T_\text{flow}$ or the number of steps, is unclear. Nevertheless, improving the proposals would have a big impact on the efficiency. As discussed earlier, at the proposal stage we estimate the behavior of the effective action as function of $x_i$ via a quadratic approximation.  This approximation gets worse for larger values of $T_\text{flow}$ as larger flow amplifies the anisotropies in the proposal space so that the flatness and steepness of the different directions become more pronounced. For our model with the particular parameters we had, a value of $T_\text{flow}=0.2$ was enough to overcome the sign problem (see Fig.~\ref{fig:histogram}), but for problems that require higher flow, as it will inevitably be the case in systems with more degrees of freedom, the proposals have to be optimized further. Secondly, the most costly part of our algorithm is the computation of the Jacobian $J$ that is performed at every update. 
Using an estimator which is cheaper and reweighting the difference at every measurement would significantly reduce the computational cost. Several such estimators have been found in the context of Lefschetz thimbles \cite{Alexandru:2016lsn} and proved to be very useful, but finding one that is applicable to our method that is flowing from $\R^N$ is still an open problem. These issues are left for future work. 

\section{Discussion and conclusions}

We have presented a new method to stochastically compute 
real time correlators of the kind required for the calculation of transport coefficients.
We pointed out that the straightforward separation of phase leads to a sign problem that is, in a sense, infinitely bad.
The method, obviously inspired by the ``Lefshetz thimble" approach \cite{Cristoforetti:2012su,Cristoforetti:2013qaa,Cristoforetti:2013wha,Fujii:2013sra,Kanazawa:2014qma,Tanizaki:2014xba,Tanizaki:2014tua,Fujii:2015bua,Fujii:2015vha,Fukushima:2015qza,Tanizaki:2015rda,DiRenzo:2015foa} \footnote{In a broader context see also \cite{Witten:2010zr,Witten:2010cx,Harlow:2011ny,Cherman:2014ofa,Behtash:2015zha,Behtash:2015kva}}, is based on a deformation of the region of integration of the path integral into complex space but, contrary to the Lefshetz thimble approach it does not require {\it a priori} knowledge of the position of the critical points, their thimbles and the contribution of each one to the original integral. We test it with success, on a simple quantum mechanical model where the complex Langevin method fails to converge.
Even though there is no theoretical obstruction to use the method in problems with larger degrees of freedom, such as field theory, at its current stage, the slow convergence of the method makes it expensive to do so.
 Future work should focus on improving the convergence rate by developing more efficient Metropolis proposals and by finding of a good estimator of the Jacobian.

\section{Acknowledgments }
We thank G. Aarts for helpful comments. A.A. is supported in part by the National Science Foundation CAREER grant PHY-1151648. 
A.A. gratefully acknowledges the hospitality 
of the Physics Department at the University of Maryland where part of this work was 
carried out.
G.B., P.F.B., S.V and N.C.W.  are supported by U.S. Department of Energy under Contract No. DE-FG02-93ER-40762.

\bibliography{realtime-MC}
\end{document}